\documentclass[12pt]{article}
\usepackage{latexsym,exscale}
\usepackage{a4,epsf}

\newcommand{\mev}{\mathrm{MeV}}

\begin{document}

\title{Polarized $\rho$ mesons and the asymmetry between $\Delta\bar{d}(x)$ and $\Delta\bar{u}(x)$ in the sea of the nucleon}

\author{R.\ J.\ Fries\thanks{email: \,rainer.fries@physik.uni-regensburg.de}
  \ and A. Sch\"afer\\ \\
  \it Institut f\"ur Theoretische Physik,\\
  \it Universit\"at Regensburg, 93040 Regensburg}

\date{}
\maketitle

\begin{abstract}
We present a calculation of the polarized $\rho$ meson cloud in a nucleon using time-ordered perturbation theory in two different variants advocated in the literature. We calculate the induced difference between the distributions $\Delta\bar{d}(x)$ and $\Delta\bar{u}(x)$. We use a recent lattice calculation to motivate an ansatz for the polarized valence quark distribution of the $\rho$ meson. Our calculations show that the two theoretical approaches give vastly different results. We conclude that $\Delta\bar{d}(x)- \Delta\bar{u}(x)$ can be of relevant size with important consequences for the combined fits of polarized distribution functions.
\end{abstract}

\medskip

\begin{figure}[t]
  \epsfxsize = 5 cm
  \centerline{\epsfbox{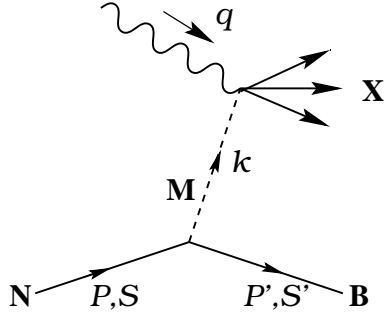}}
  \caption{The $MNB$ process. A nucleon $N$ splits up into a meson and a baryon
    and the meson is hit by a photon in a deep inelastic scattering 
    experiment.
    \label{mnb}}
\end{figure}

It is experimentally established beyond doubt that $\bar{d}(x)$ and $\bar{u}(x)$ in the nucleon differ substantially \cite{nmc91,nmc94,e866}. 
The difference can be described as due to the meson cloud in the nucleon. The relevant process is shown in fig. \ref{mnb}. The nucleon splits up into a meson and a baryon and the meson is hit by the photon in a deep inelastic scattering experiment. 

The dominant role in the unpolarized case is played by the pion. It is clear that a proton prefers being accompanied by a $\pi^+$ instead of an $\pi^-$ since in the first case the remaining baryon can be a neutron, but in the second case the remaining baryon with the lowest lying mass is the $\Delta(1232)$ resonance.
Therefore one can find an enhancement of $\bar{d}$ quarks over $\bar{u}$ quarks in the proton. 
A large number of papers has been published on this subject, e.g.~\cite{hemi90,kumlond,hwang91,melni93,fra95,holtmann,kumano97}.

Proceeding a step further in this direction one can ask the question whether there is also an asymmetry in the polarized sea, more precisely between the distributions $\Delta\bar{d}(x)$ and $\Delta\bar{u}(x)$.
Up to now it is not possible to extract such an asymmetry from experiments but the situation could change in the future.
It is clear that the meson cloud model predicts such an asymmetry. The mechanism is the same as in the unpolarized case, now with the lightest polarizable meson --- the $\rho$-Meson with a mass of $770~\mev$ --- in the role of the pion. 
There have already been attempts to derive this asymmetry from quark distribution functions calculated in the chiral quark-soliton model \cite{diak96,diak97}.

In this publication we present a calculation of the difference between the light polarized antiquark distributions.  
We take into account the processes where a nucleon $N$ fluctuates into $\rho$ and $N$ ($\rho NN$) and into $\rho$ and $\Delta$ ($\rho N\Delta$).
Let us consider the process in fig. \ref{mnb} in an infinite momentum frame with $P$, $P'$, $k$ and $q$ being the four-momenta of the incoming nucleon, the outgoing baryon, the meson and the photon respectively. We are only interested in spin 1 mesons and $S$, $S'$ and $\lambda$ shall denote the helicities of the nucleon, the baryon and the meson.
 
The contribution $\delta \bar{q}^\sigma$ of the $MNB$ process in fig.~\ref{mnb} to the distribution of antiquarks of a given flavour with helicity $\sigma$ in a nucleon $N$ is described by the well known convolution formula \cite{holtmann}
\begin{equation}
  x \delta \bar{q}^\sigma(x) = \sum_{\lambda}\int_x^1 dy f_{MNB}^\lambda(y) 
  \frac{x}{y} 
  \bar{q}^\sigma_{M,\lambda} \left( \frac{x}{y} \right) 
  \label{conv}
\end{equation}
where 
\begin{equation}
  x = \frac{Q^2}{2 P\cdot q}
\end{equation}
is the usual Bjorken variable with $Q^2 = -q^2$ and 
\begin{equation}
  y= \frac{k\cdot q}{P \cdot q}
\end{equation}
is the momentum fraction of the nucleon carried by the meson. $f_{MNB}^\lambda(y)$ denotes the probability distribution for a meson $M$ with helicity $\lambda$ in the nucleon due to the $MNB$ process and $\bar{q}^\sigma_{M,\lambda}$ gives the probability to find an antiquark of a given flavour with helicity $\sigma$ in a Meson with helicity $\lambda$.
As a shorthand notation for the convolution we use $\delta\bar{q}^\sigma = f^\lambda_{MNB} \otimes \bar{q}^\sigma_{M,\lambda}$ in the following. 
Now let us consider a proton with helicity +1/2 (or briefly $\uparrow$) as the incoming nucleon.
With the usual definitions of the helicity asymmetry in the nucleon
\begin{equation}
  \Delta \bar{q} = \bar{q}^\uparrow - \bar{q}^\downarrow
\end{equation}
and in the meson
\begin{equation}
  \Delta \bar{q}_{M} = \Delta\bar{q}_{M,1}^\uparrow - \Delta\bar{q}_{M,
   1}^\downarrow = \Delta\bar{q}_{M,-1}^\downarrow - \Delta\bar{q}_{M,
   -1}^\uparrow
\end{equation}
and with the notation
\begin{equation} 
  \Delta f_{MNB} = f^1_{MNB} - f^{-1}_{MNB}
\end{equation}
for the polarization of the meson cloud
we can write e.g.~the contribution of the process $p \to \rho^+ + n$ to the polarization of the $\bar{d}$ quarks as
\begin{equation}
  \delta (\Delta\bar{d}) = \Delta f_{\rho^+ pn} \otimes\Delta \bar{d}_{\rho^+}.
\end{equation}
Charge symmetry tells us that 
\begin{equation}
  \Delta\bar{d}_{\rho^+} = \Delta\bar{u}_{\rho^-} = 2 \Delta\bar{d}_{\rho^0}
   = 2 \Delta\bar{u}_{\rho^0} =: \Delta v_\rho.
\end{equation}
Taking into account all possible isospin channels of the $\rho NN$ and $\rho N\Delta$ processes we obtain
\begin{eqnarray}
  \delta (\Delta \bar{d}) &=& \left( \frac{5}{6}\Delta f_{\rho NN} +\frac{1}{3}
  \Delta f_{\rho N\Delta} \right) \otimes \Delta v_\rho   \\
  \delta (\Delta \bar{u}) &=& \left( \frac{1}{6}\Delta f_{\rho NN} +\frac{2}{3}
  \Delta f_{\rho N\Delta} \right) \otimes \Delta v_\rho
\end{eqnarray}
and finally
\begin{equation}
  \Delta\bar{d}-\Delta\bar{u} = \left( \frac{2}{3} \Delta f_{\rho NN} 
  -\frac{1}{3} \Delta f_{\rho N\Delta} \right) \otimes \Delta v_\rho.
\end{equation}
$f_{\rho NN}$ and $f_{\rho N\Delta}$ are the total $\rho$ distributions summed over all isospin channels with a proton as the initial nucleon.

For the calculation of the $\rho$ distributions we use time-ordered perturbation theory (TOPT) in the infinite-momentum frame \cite{weinb66,drell70} which guarantees on-shellness for the mesons in contrast to the usual covariant approach where the meson is not on its mass shell. The TOPT approach has become very popular in recent times since the fact that the meson is no longer off-shell removes a big problem of the early calculations. As a disadvantage there is no energy conserving $\delta$-function at the $MNB$ vertex.

In the infinite momentum frame the three momentum of the proton is $(0,0,p)$ with $p$ tending to infinity and the momentum of the meson can then be written as
$(k_\perp \cos\varphi, k_\perp \sin\varphi, yp)$
with a transverse momentum $k_\perp$. It has been shown that only time-ordered diagrams with all particles moving forward in the infinite momentum frame contribute, others are suppressed by $1/p^2$. Therefore $y$ takes only values between 0 and 1 \cite{melni93,holtmann,drell70}.
One can write the meson distribution as an integral
\begin{equation}
  f^\lambda_{MNB} (y) = \sum_{s'} \int_0^\infty dk_\perp^2
  {\left | \Phi^{MB}_{\lambda s'}(y,k_\perp^2) \right|}^2
\end{equation}
where $\Phi^{MB}_{\lambda,s'}$ is the helicity dependent amplitude for the $MNB$ process considered. The helicity $s'$ of the baryon is not observed, so it is summed over.
The amplitudes are given by \cite{holtmann}
\begin{equation}
  \Phi^{MB}_{\lambda s'}(y,k_\perp^2) = \frac{\sqrt{MM'}}{2\pi \sqrt{y(1-y)}}
  \frac{V^{MB}_{\lambda s'}(y,k_\perp^2)G(y,k_\perp^2)}{M^2-s^2_{MB}(y,
  k_\perp^2)}.
\end{equation}
The masses of the nucleon, the outgoing baryon and the $\rho$ meson are denoted here with $M$, $M'$ and $m$ and 
\begin{equation}
  s_{MB}^2(y,k_\perp^2) = \frac{k_\perp^2 +M^2}{1-y}+\frac{k_\perp^2+m^2}{y}
\end{equation}
is the invariant squared mass of the meson-baryon state. The vertex function
$V^{MB}_{\lambda s'}$ contains the vector and spinor structure of the vertex contracted with the fields and depends on the Lagrangian used for the effective meson nucleon interaction. As usual a form factor must be introduced for the interaction which is parameterized in this case in exponential form 
\begin{equation}
  G(y,k_\perp^2) = \exp \left[ \frac{M^2-s_{MB}^2(y,k_\perp^2)}{2 
  \Lambda_{MNB}^2} \right]
\end{equation}
as a function of the squared mass $s_{MB}$ and with the cutoff parameter $\Lambda_{MNB}$.

The Lagrangians of the effective $\rho NN$ and $\rho N\Delta$ interaction are \cite{bonn}
\begin{equation}
  \mathcal{L}_{\rho NN} = g_{\rho NN} \bar{\psi} \gamma^\mu \phi_\mu \psi
  + f_{\rho NN} \bar{\psi} \sigma^{\mu\nu} \psi (\partial_\mu \phi_\nu -
  \partial_\nu \phi_\mu),
\end{equation}
and
\begin{equation}
  \mathcal{L}_{\rho N\Delta} = g_{\rho N\Delta} \bar{\psi} i\gamma_5 \gamma^\mu \Psi^\nu
  (\partial_\mu \phi_\nu - \partial_\nu \phi_\mu) + \mathrm{h.c.}
\end{equation}
with the Dirac field $\psi$ for the nucleon, the Rarita-Schwinger field $\Psi^\nu$ for the $\Delta$ and the vector field $\phi^\mu$ for the $\rho$ meson.
The vertex functions therefore are
\begin{eqnarray}
  V^{\rho N}_{\lambda S'} &=& g_{\rho NN}\bar{u}(P',s') \gamma^\mu 
  \epsilon^*_\mu (k,\lambda) u(P,1/2) + \nonumber \\
  && {}+ 2 f_{\rho NN} \bar{u}(P',S')
  i\sigma^{\mu\nu} k_\mu \epsilon^*_\nu u(P,1/2), \label{vert1}\\
  V^{\rho \Delta}_{\lambda s'} &=& g_{\rho N\Delta} \bar{U}^\nu (P',S')
  i \gamma_5 \gamma^\mu u(p,1/2) \left[ i k_\mu \epsilon^*_\nu (k,\lambda)
  - i k_\nu \epsilon^*_\mu(k,\lambda) \right], \label{vert2}
\end{eqnarray}
where $u$ is a Dirac spinor, $U^\mu$ a Rarita-Schwinger spinor and $\epsilon^\mu$ the polarization vector of the $\rho$ meson.

\begin{figure}[t]
  \epsfxsize = 10cm
  \centerline{\epsfbox{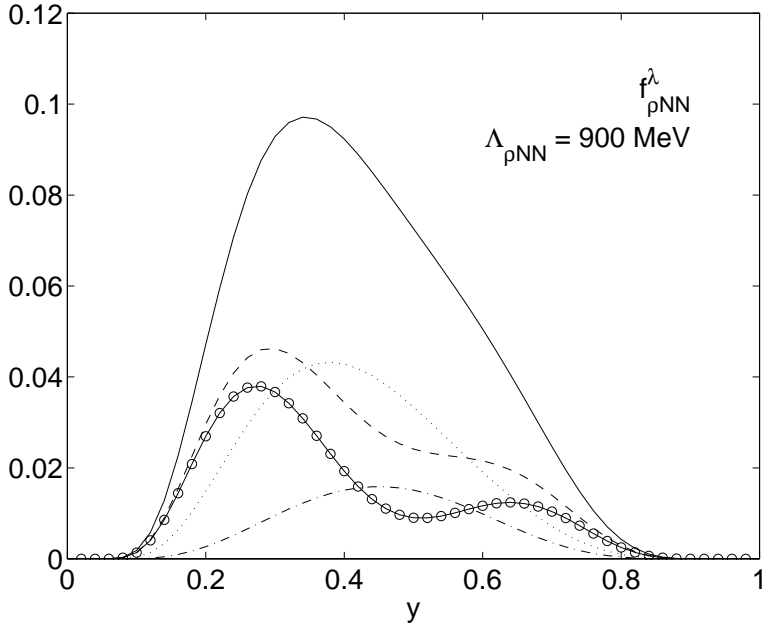}}
  \caption{Distributions $f^\lambda_{\rho NN} (y)$ of the $\rho$ mesons from 
    the $\rho NN$ process for helicities $\lambda=1$ (dashed line), 0 (dotted 
    line) and $-1$ (dashed-dotted line). The solid line shows the sum over all
    helicities and the line with circles the polarization $\Delta f_{\rho NN}
    (y) = f^1_{\rho NN}(y) - f^{-1}_{\rho NN}(y)$. Method (A) and a cutoff 
    $\Lambda_{\rho NN} = 900~\mev$ in the formfactor are used.
    \label{rnnvert}}
\end{figure}

\begin{figure}[h]
  \epsfxsize = 10cm
  \centerline{\epsfbox{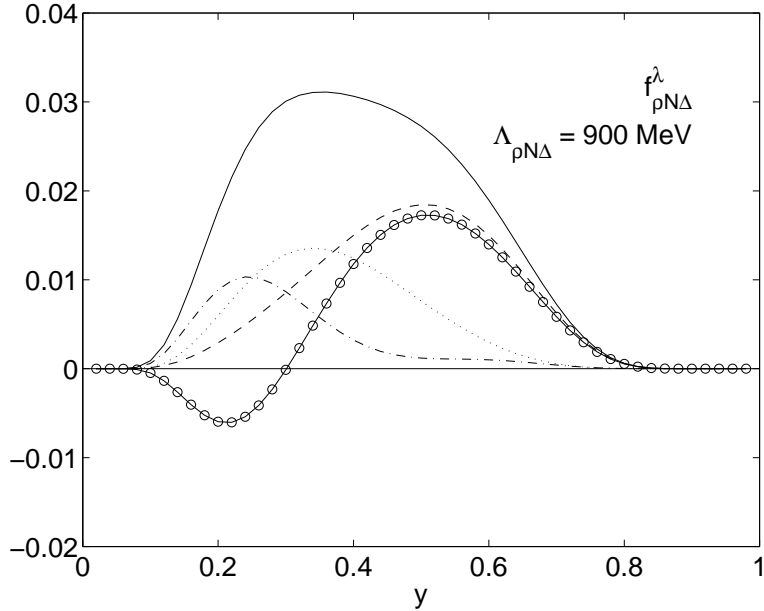}}
  \caption{The same as in fig.~\ref{rnnvert}, now for the $\rho N\Delta$
    process obtained from method (A) and with a cutoff $\Lambda_{\rho N\Delta}
    = 900~\mev$.
  \label{rndvert}}
\end{figure}

\begin{figure}[p]
  \epsfxsize = 10cm
  \centerline{\epsfbox{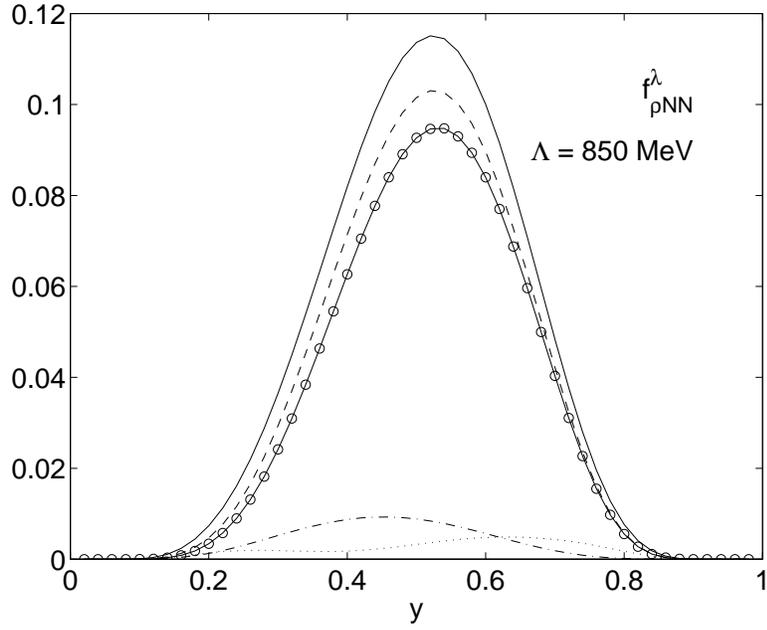}}
  \caption{The same as fig.~\ref{rnnvert}, now obtained from method (B) with 
    a cutoff parameter $\Lambda_{\rho NN} = 850~\mev$. 
  \label{ecrnnvert}}
\end{figure}

\begin{figure}[p]
  \epsfxsize = 10cm
  \centerline{\epsfbox{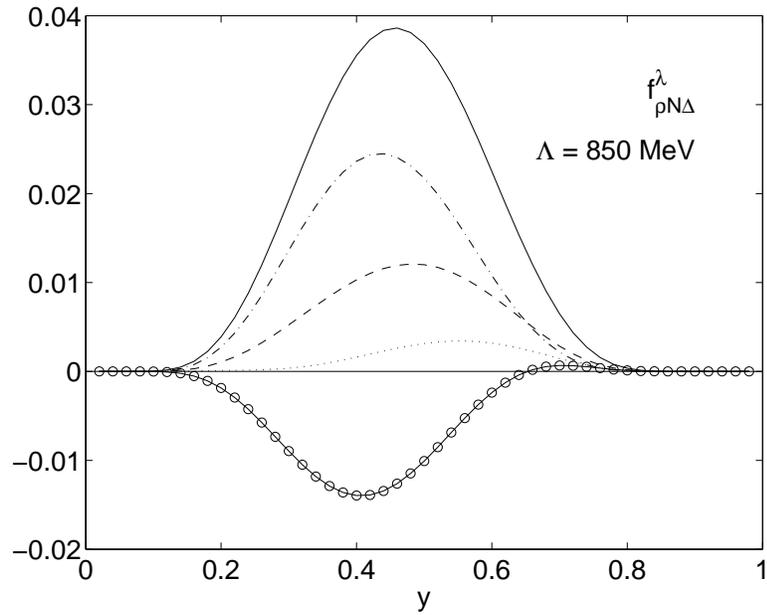}}
  \caption{The same as fig.~\ref{rndvert}, now obtained from method (B) with 
    a cutoff parameter $\Lambda_{\rho N\Delta} = 850~\mev$.
  \label{ecrndvert}}
\end{figure}

There were some discussions whether this is the correct choice \cite{holtmann,thomas93}. The reason for this is the appearence of some explicit factors of $k$ in (\ref{vert1}) and (\ref{vert2}). They arise from the derivatives of the meson field in the interaction Lagrangians. 
The argument is that in spite of the absent energy conservation one should not use the on-shell momentum vector $k$ of the meson in that case (A) but replace it by the vector $\tilde k = P-P'$ instead (B). Note that only the space-like components of both vectors coincide. The definition of $y$ and the polarization vectors of the meson remain unchanged.
We do not want to make a decision in this matter. Obviously such calculations still lack a really solid theoretical basis and have to be interpreted as effective models to be judged by their phenomenological success. 
Both methods have been used to calculate unpolarized contributions of the $\rho$ meson to quark distributions, see e.g.\ \cite{melni93} for (A) and \cite{holtmann,thomas93} for (B). In this respect both procedures do roughly equally well.

We will present here calculations for both methods to get a feeling for the theoretical uncertainties involved.
We find in fact large differences between the numerical results indicating that the theoretical treatment needs substantial improvement.
Combining all the input we obtain for the helicity dependent and isospin summed distribution functions from the $\rho NN$ process the results given in the appendix for both prescriptions.
For the numerical calculations we used the coupling constants $g_{\rho NN}^2 /
(4\pi) = 0.84$, $f_{\rho NN} = 6.1 g_{\rho NN}/(4M)$ and $g_{\rho N\Delta}^2/(4\pi) = 20.45$ given in \cite{holtmann,bonn}. Figs.~\ref{rnnvert} and \ref{rndvert} show the distributions of the $\rho$ mesons from the $\rho NN$ and the $\rho N\Delta$ process respectively for different helicities obtained from method (A). In Figs.~\ref{ecrnnvert} and \ref{ecrndvert} the same curves are plotted from method (B).

Note that method (A) favours the $\rho$ mesons with helicity $+1$ without suppressing the other helicities too much. In contrast, method (B) leads to an overwhelming dominance of helicity $+1$ for the $\rho NN$ channels but for $\rho N\Delta$ we have more mesons with helicity $-1$. 
Note also that method (B) gives a peak at $x\approx 0.55$ for the unpolarized distribution from the $\rho NN$ process while the distributions calculated with method (A) peak below $0.5$. 
From the mass ratio of the $\rho$ on one side and the nucleon or the $\Delta$ on the other side we would naively expect the meson to carry less than 50\% of the total momentum in average.

\begin{figure}[t]
  \epsfxsize = 10cm
  \centerline{\epsfbox{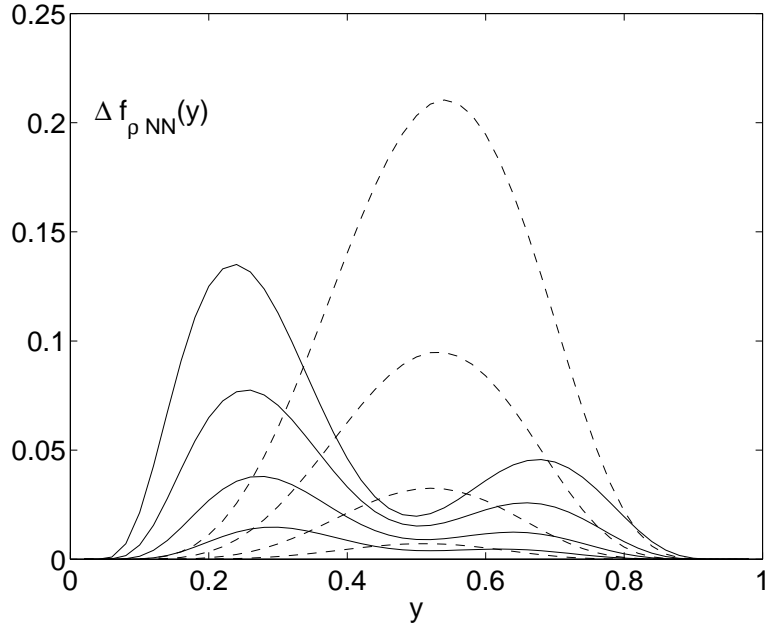}}
  \caption{The polarization $\Delta f_{\rho NN}(y)$ of the $\rho$ meson cloud 
    from 
    the $\rho NN$ process calculated with method (A) and cutoff parameters
    $\Lambda_{\rho NN} = 1100~\mev$, $1000~\mev$, $900~\mev$ and $800~\mev$ 
    (solid lines from top to bottom) and with method (B) and
    the cutoffs $\Lambda_{\rho NN} = 950~\mev$, $850~\mev$, $750~\mev$ and 
    $650~\mev$ (dashed lines from top to bottom).
    \label{drnncomp}}
\end{figure}

\begin{figure}[h]
  \epsfxsize = 10cm
  \centerline{\epsfbox{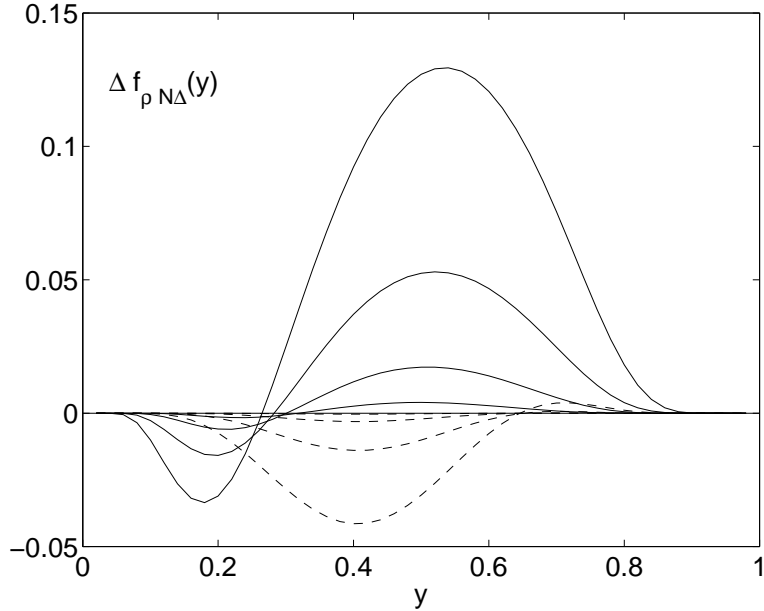}}
  \caption{The same as fig.~\ref{drnncomp} now for the $\rho N\Delta$ 
    process and again cutoff parameters $\Lambda_{\rho N\Delta}=1100~\mev$, 
    $1000~\mev$, $900~\mev$ and $800~\mev$ (solid lines from top to bottom) for    method (A) and $\Lambda_{\rho N\Delta}=950~\mev$, 
    $850~\mev$, $750~\mev$ and $650~\mev$ (dashed lines from bottom to top 
    at $y=0.4$) for method (B).
    \label{drndcomp}}
\end{figure}

In Figs.~\ref{drnncomp} and \ref{drndcomp} we compare the polarized distributions $\Delta f(y)$, which are relevant for our problem, for both methods and different cutoffs. The cutoffs for the $\rho NN$ and $\rho N\Delta$ vertices are not very well known. Holtmann \textit{et al.}\ \cite{holtmann} use $\Lambda = 1100~\mev$ for the $\rho NN$ vertex and $\Lambda = 980~\mev$ for the the $\rho N\Delta$ vertex from fits to diffractive scattering data. With view to the serious problems of fixing the $\pi NN$ formfactor \cite{fra95,fries} we decided to vary the cutoff parameters and keep the unpolarized $\rho$ distributions of both methods in comparable size. The unpolarized curves can also be compared with those in \cite{melni93} for method (A) and in \cite{thomas93} for method (B). 
Unfortunately the dependence on the cutoff is for the $\rho$ much larger than for the pion as already stated by Thomas and Melnitchouk \cite{melni93}.
Figs.~\ref{drnncomp} and \ref{drndcomp} show the large differences between (A) and (B).
One notes also that both models predict a partially negative polarization of the $\rho$ cloud from the $\rho N\Delta$ process.

A big obstacle for considering contributions of mesons to the polarized sea of the nucleon is up to now our lack of knowledge about the polarized structure functions of vector mesons \cite{hood}. It is common to set the unpolarized valence quark distribution of the $\rho$ equal to that of the pion. For our purpose we need the helicity asymmetry $\Delta v_\rho(x)$ of a valence quark in the $\rho$ meson which has not yet been measured. 
A recent lattice calculation \cite{best} shows that the polarization of the valence quarks in the $\rho$ should be approximately 60\%. Therefore we set $\Delta v_\rho(x)  = 0.6 \, v_\pi(x)$ where $v_\pi$ is the valence quark distribution of the pion. Here we use the parameterizations of Gl\"uck, Reya and Vogt for the pion \cite{glueck}. 
This is a rather bold ansatz but it enables us to get a first estimate for the flavour asymmetry in the polarized sea of the nucleon.

\begin{figure}[h]
  \epsfxsize = 10cm
  \centerline{\epsfbox{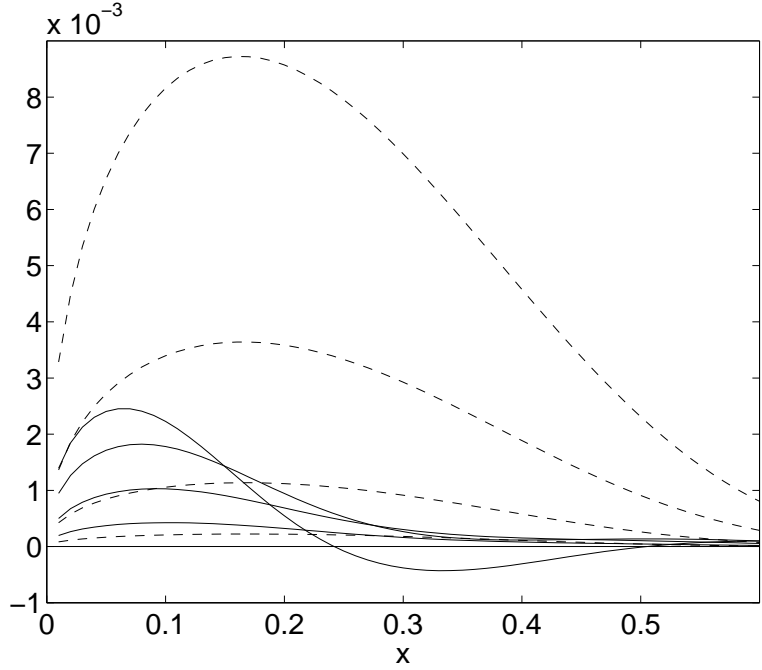}}
  \caption{The asymmetry $x(\Delta\bar{d}(x)-\Delta\bar{u}(x))$ calculated with
    the $\rho$ distributions of method (A) and cutoff parameters
    $\Lambda_{\rho NN} = \Lambda_{\rho N\Delta} = 1100~\mev$, $1000~\mev$, 
    $900~\mev$ and $800~\mev$ (solid lines from top to bottom) and with method
    (B) and $\Lambda_{\rho NN} = \Lambda_{\rho N\Delta} = 950~\mev$, 
    $850~\mev$, $750~\mev$ and $650~\mev$ (dashed lines from top to bottom) at
    $Q^2=2~\mathrm{GeV}^2$.  
    \label{dd-du1}}
\end{figure}

\begin{table}
\centerline{
\begin{tabular}{|c|c||c|c|}
  \hline
  \multicolumn{2}{|c||}{\rule[-3mm]{0mm}{8mm}(A)} & 
  \multicolumn{2}{c|}{(B)} \\ \hline
  $I$    & $\Lambda [\mev]$ &  $I$ & $\Lambda [\mev]$ \\ \hline\hline
  0.0064 & 1100 & 0.027   & 950 \\
  0.0052 & 1000 & 0.011   & 850 \\
  0.0030 & 900  & 0.0034  & 750 \\
  0.0013 & 800  & 0.00067 & 650 \\ \hline
\end{tabular} }
  \caption{Results for the Integral $I=\int_{0.01}^1dx\, (\Delta \bar{d}(x)-
  \Delta \bar{u}(x) )$ for both methods and different formfactors $\Lambda=
  \Lambda_{\rho NN} = \Lambda_{\rho N\Delta}$.
  \label{table}}
\end{table}

In figs.~\ref{dd-du1} we present our final results. The contributions to 
$x\Delta\bar{d}(x)-x\Delta\bar{u}(x)$ from the $\rho NN$ and $\rho N\Delta$ processes are calculated with prescriptions (A) and (B). For both methods we chose four different cutoff parameters $\Lambda = \Lambda_{\rho NN}=\Lambda_{\rho N\Delta}$. 
The first point which attracts attention is the large difference between the results of (A) and (B). 
In view of the wide range of possible results it is clear that the influence of the polarized $\rho$ mesons is not necessarily negligible. Current parameterizations for $x\Delta\bar{q}(x)$ give typically a value around 0.02 or less for $x=0.1$ \cite{gehr95,glueck95}. 
Thus the asymmetry in the polarized nucleon sea could be a 50\% effect. 
References \cite{diak96,diak97} predict even a larger asymmetry.
Note that in these papers $\Delta \bar{u}(x)$ and $\Delta \bar{d}(x)$ are defined with the opposite sign \cite{polyakov}.
Tab.~\ref{table} gives our numerical values for
\begin{equation}
  I = \int_{0.01}^1 dx\, \left( \Delta \bar{d}-\Delta\bar{u} \right)
\end{equation}
for the different models plotted in fig.~\ref{dd-du1}.

Let us summarize our work. We calculated the helicity dependent distributions of $\rho$ mesons in a proton with two models.  
Under convolution the different shapes of the $\rho$ distributions do not translate directly into the quantity $x\Delta\bar{d}(x)-x\Delta\bar{u}(x)$ but remarkable differences remain.
Also the determination of the correct formfactors from unpolarized data should be improved.
Bearing all that caveats in mind it is clear that our results show that the flavour asymmetry in the polarized sea can be relevant for the present DESY, SLAC and CERN spin experiments.

We strongly thank G. Piller, W. Weise and M. G\"ockeler for discussions. The fruitful conversation with A. W. Thomas, W. Melnitchouk and A. Szczurek is gratefully acknowledged. This work was supported by BMBF.

\begin{appendix}

\section*{Appendix}

Here we give our results for the $\rho$ distributions. Using the on-shell meson momentum $k$ at the vertex (A) we obtain
\renewcommand{\arraystretch}{1.5}
\begin{eqnarray}
  f^1_{\rho NN} (y) &=& \frac{3}{8\pi^2 y^3(1-y)^2} \int_0^\infty 
  \frac{d k_\perp^2 \, F^2(y,k_\perp^2)}{{(M^2-s^2_{\rho N}(y,k_\perp^2))}^2}
  \nonumber \\
  && \Big\{ g^2 \left[ k_\perp^2 + y^4 M^2 \right] +  
  +4 gfyM \left[ k_\perp^2 + y\left( y^2M^2- (1-y) m^2\right) \right]  + 
  \nonumber \\
  && {}+ 4f^2\left[ y^2M^2 k_\perp^2+ {\left( y^2M^2-(1-y)m^2 \right)}^2 
  \right] \Big\}, \\ \nonumber \\
  f^0_{\rho NN} (y) &=& \frac{3}{16\pi^2 y^3(1-y)^2m^2} \int_0^\infty 
  \frac{d k_\perp^2 \, F^2(y,k_\perp^2)}{{(M^2-s^2_{\rho N}(y,k_\perp^2))}^2}
  \nonumber \\
  && \Big\{ g^2 {\left[ k_\perp^2 + y^2M^2-(1-y)m^2 \right]}^2 + 4gfy^2m^2M
  \big[ k_\perp^2+y^2M^2-  \nonumber \\
  && {} -(1-y)m^2 \big] + 4f^2m^4 \left[ (2-y)^2 k_\perp^2 +y^4 M^2 
  \right] \Big\},
  \\ \nonumber \\
  f^{-1}_{\rho NN} (y) &=& \frac{3}{8\pi^2 y^3(1-y)^2} \int_0^\infty 
  \frac{d k_\perp^2 \, F^2(y,k_\perp^2)}{{(M^2-s^2_{\rho N}(y,k_\perp^2))}^2}
  \nonumber \\
  && k_\perp^2\Big\{ g^2 (1-y)^2 -4fgy(1-y)M+ 4f^2 
  \left[ k_\perp^2 +y^2M^2 \right]
  \Big\}
\end{eqnarray}
for the $\rho NN$ process and
\begin{eqnarray}
  f^1_{\rho N\Delta}(y) &=& 
  \frac{g^2}{24\pi^2 y^3 (1-y)^4 M'^2} \int_0^\infty \frac{d k_\perp^2 \,
  F^2(y,k_\perp^2)}{{(M^2-s^2_{\rho \Delta}(y,k_\perp^2))}}
  \nonumber \\
  && \Big\{ k_\perp^6 + k_\perp^4 M'^2
  \left[ 3-4y(1-y) \right] + \nonumber \\
  && {}+ k_\perp^2 M' \left[ 4y(1-y)^3m^2M + 2y^4M'^3 + y^2{(2-y)}^2 M'^3
  \right] - \nonumber \\
  && {}- 2y^2(1-y)^3 m^2 MM'^3 +y^4 M'^6 + (1-y)^6 m^4 M^2 \Big\},
  \\ \nonumber \\
  f^0_{\rho N\Delta}(y) &=& 
  \frac{g^2 m^2}{12\pi^2 y^3 (1-y)^2 M'^2} \int_0^\infty \frac{d k_\perp^2 \,
  F^2(y,k_\perp^2)}{{(M^2-s^2_{\rho \Delta}(y,k_\perp^2))}^2}
  \nonumber \\
  && \Big\{ k_\perp^4 +k_\perp^2 \left[ 2M'^2+
  (1-y)^2(M^2+M'^2) \right] + \nonumber \\ 
  && {}+ y^2 M'^4 -2y^2(1-y)MM'^3 + y^2(1-y)^2 M^2M'^2 \Big\},
  \\ \nonumber \\
  f^{-1}_{\rho N\Delta}(y) &=& 
  \frac{g^2}{24\pi^2 y^3 (1-y)^2 M'^2} \int_0^\infty \frac{d k_\perp^2 \,
  F^2(y,k_\perp^2)}{{(M^2-s^2_{\rho \Delta}(y,k_\perp^2))}^2}
  \nonumber \\  
  && \Big\{ k_\perp^4 M^2 + k_\perp^2 \left[
  4y^2 M^2M'^2 - 4y(1-y)m^2MM' +(1-y)^2m^4 \right] + \nonumber \\ 
  && {}+ 3y^4M^2M'^4 -6y^2(1-y)m^2MM'^3 +
  3(1-y)^2m^4M'^2 \Big\}, 
\end{eqnarray}
for the $\rho N\Delta$ process.

For the vector $\tilde k=P-P'$ (B) the results are
{
\renewcommand{\arraystretch}{1.5}
\begin{eqnarray}
  f^1_{\rho NN} (y) &=& \frac{3}{8\pi^2 y^3(1-y)^2} \int_0^\infty 
  \frac{d k_\perp^2 \, F^2(y,k_\perp^2)}{{(M^2-s^2_{\rho N}(y,k_\perp^2))}^2}
  \nonumber \\
  && \Big\{ g^2 \left[ k_\perp^2 + y^4 M^2 \right] +  
  +4 gfM \left[ y(1+y)k_\perp^2 + 2y^4M^2 \right]  + 
  \nonumber \\
  && {}+ 4f^2\left[ k_\perp^4 + 5k_\perp^2 y^2M^2 +4y^4M^4
  \right] \Big\}, \\ \nonumber \\
  f^0_{\rho NN} (y) &=& \frac{3}{16\pi^2 y^3(1-y)^4m^2} \int_0^\infty 
  \frac{d k_\perp^2 \, F^2(y,k_\perp^2)}{{(M^2-s^2_{\rho N}(y,k_\perp^2))}^2}
  \nonumber \\
  && {\left[ k_\perp^2 + y^2M^2-(1-y)m^2 \right]}^2 \nonumber \\
  && \left[ g^2(1-y)^2 -2gfy^2(1-y)M +f^2 \left( k_\perp^2(2-y)^2
  +y^4M^2 \right) \right],
  \\ \nonumber \\
  f^{-1}_{\rho NN} (y) &=& \frac{3}{8\pi^2 y^3(1-y)^2} \int_0^\infty 
  \frac{d k_\perp^2 \, F^2(y,k_\perp^2)}{{(M^2-s^2_{\rho N}(y,k_\perp^2))}^2}
  \nonumber \\
  && k_\perp^2 \Big\{ g^2 (1-y)^2 -4fgy(1-y)M + 4 f^2 \left[ k_\perp^2 
  +y^2M^2 \right] \Big\}, \\ \nonumber \\
  f^1_{\rho N\Delta}(y) &=& \frac{g^2}{24\pi^2 y^3(1-y)^4M'^2} \int_0^\infty 
  \frac{d k_\perp^2 \, F^2(y,k_\perp^2)}{{(M^2-s^2_{\rho \Delta}(y,
  k_\perp^2))}^2} \nonumber \\
  && \Big\{ k_\perp^6 +k_\perp^4 \big[ (1-y)^4 M^2 - 4(1-y)^2yMM' + 
  \nonumber \\ 
  && {} +(4y^2-4y+3)M'^2 \big] + k_\perp^2 \big[ 2y^4M'^4 
     + y^2{(2-y)}^2 M'^4 - \nonumber \\
  && {} -2(1-y)^5y M^4 +4(1-y)^3y^2M^3M' -2(1-y)^2y^2MM'^3 + 
  \nonumber \\ 
  && {}+ 2(1-y)^4y M^2M'^2 \big] + (1-y)^6y^2M^6 -2(1-y)^5y^2M^4M'^2- 
  \nonumber \\
  && {} -2(1-y)^3y^3M^3M'^3 +(1-y)^4y^2M^2M'^4+ \nonumber \\
  && {}+ 2(1-y)^2y^3MM'^5+ y^4M'^6  \Big\}, \\ \nonumber \\ 
  f^0_{\rho N\Delta}(y) &=& \frac{g^2}{48\pi^2 y^3(1-y)^4m^2M'^2} 
  \int_0^\infty 
  \frac{d k_\perp^2 \, F^2(y,k_\perp^2)}{{(M^2-s^2_{\rho \Delta}(y,
  k_\perp^2))}^2} \nonumber \\
  && \Big\{ k_\perp^8 + k_\perp^6 \big[ (y^2+3)M'^2+(1-y)(1-3y)M^2- 
  \nonumber \\ 
  && {}-2(1-y)m^2 \big] + k_\perp^4 \big[ {\big( ((1-y)M^2-M'^2)y 
  +(1-y)m^2 \big)}^2 - \nonumber \\
  && {} -2 \big( ((1-y)M^2-M'^2)y +(1-y)m^2 \big)   
  \nonumber \\
  && {} \big( (1-y)^2M^2+(2+(1-y)^2)M'^2 \big)+ y^2 M'^4 - \nonumber \\
  && {} -2(1-y)y^2MM'^3 + (1-y)^2y^2M^2M'^2 \big]+ 
  \nonumber \\
  && {}+ k_\perp^2 \big[ {\big( ((1-y)M^2-M'^2)y +(1-y)m^2 \big)}^2
  \nonumber \\
  && \big( (1-y)^2M^2+(2+(1-y)^2)M'^2 \big) - \nonumber \\
  && {}- 2 \big( ((1-y)M^2-M'^2)y +(1-y)m^2 \big)\big( y^2M'^4 - 
  2(1-y)y^2 MM'^3 + \nonumber \\
  && {} +(1-y)^2y^2M^2M'^2 \big) \big] + {\big[ ((1-y)M^2-M'^2)y 
  +(1-y)m^2 \big]}^2 \nonumber \\
  && \big[ y^2M'^4 - 2(1-y)y^2 MM'^3 +(1-y)^2y^2M^2M'^2 
  \big] \Big\}, \\ \nonumber \\
  f^{-1}_{\rho N\Delta}(y) &=& \frac{g^2}{24\pi^2 y^2(1-y)^3M'^2} 
  \int_0^\infty \frac{d k_\perp^2 \, F^2(y,k_\perp^2)}{{(M^2-s^2_{\rho \Delta}
  (y,k_\perp^2))}^2} \nonumber \\
  && \Big\{ k_\perp^6 +k_\perp^4 \big[ 4yMM' + y^2M^2 +(1-y)^2M^2 + 
  \nonumber \\
  && {}+ 3M'^2 +2yM'^2 \big] + k_\perp^2 \big[ (1-y)^2y^2M^4
  -4(1-y)y^2M^3M' + \nonumber \\
  && {}+ (y^2+6y)M'^4 - 2(-y^3-4y^2+3y)M^2 M'^2 + \nonumber \\
  && {}+ 10y^2MM'^3 \big] +3y^2M'^6 + 6y^3 MM'^5 + \nonumber \\  
  && {}+ 3y^4M^2M'^4 - 6(1-y)y^3M^3M'^3 + 3(1-y)^2y^2M^4 M'^2
  - \nonumber \\
  && {}+  6(1-y)y^2M^2M'^4 \Big\}.
\end{eqnarray} }

\end{appendix}

\end{document}